\overfullrule=0pt
%
%
%
%
%
%
\def\unredoffs{} 

%
%
%
%
\newbox\leftpage \newdimen\fullhsize \newdimen\hstitle \newdimen\hsbody
\tolerance=1000\hfuzz=2pt
\catcode`\@=11 
\def\bigans{b }
%
\magnification=1200\unredoffs\baselineskip=16pt plus 2pt minus 1pt
\hsbody=\hsize \hstitle=\hsize 
%
%
%
\newcount\yearltd\yearltd=\year\advance\yearltd by -1900

\def\Title#1#2{\nopagenumbers\abstractfont\hsize=\hstitle\rightline{#1}%
\vskip 1in\centerline{\titlefont #2}\abstractfont\vskip .5in\pageno=0}
\def\Date#1{\vfill\leftline{#1}\tenpoint\supereject\global\hsize=\hsbody%
\footline={\hss\tenrm\folio\hss}}
%

\def\draftmode{\message{ DRAFTMODE }\def\draftdate{{\rm preliminary draft:
\number\month/\number\day/\number\yearltd\ \ \hourmin}}%
\headline={\hfil\draftdate}\writelabels\baselineskip=20pt plus 2pt minus 2pt
 {\count255=\time\divide\count255 by 60 \xdef\hourmin{\number\count255}
  \multiply\count255 by-60\advance\count255 by\time
  \xdef\hourmin{\hourmin:\ifnum\count255<10 0\fi\the\count255}}}
\def\nolabels{\def\wrlabeL##1{}\def\eqlabeL##1{}\def\reflabeL##1{}}
\def\writelabels{\def\wrlabeL##1{\leavevmode\vadjust{\rlap{\smash%
{\line{{\escapechar=` \hfill\rlap{\sevenrm\hskip.03in\string##1}}}}}}}%
\def\eqlabeL##1{{\escapechar-1\rlap{\sevenrm\hskip.05in\string##1}}}%
\def\reflabeL##1{\noexpand\llap{\noexpand\sevenrm\string\string\string##1}}}
\nolabels
%
\global\newcount\secno \global\secno=0
\global\newcount\meqno \global\meqno=1
\def\newsec#1{\global\advance\secno by1\message{(\the\secno. #1)}
\global\subsecno=0\eqnres@t\noindent{\bf\the\secno. #1}
\writetoca{{\secsym} {#1}}\par\nobreak\medskip\nobreak}
\def\eqnres@t{\xdef\secsym{\the\secno.}\global\meqno=1\bigbreak\bigskip}
\def\sequentialequations{\def\eqnres@t{\bigbreak}}\xdef\secsym{}
\global\newcount\subsecno \global\subsecno=0
\def\subsec#1{\global\advance\subsecno by1\message{(\secsym\the\subsecno. #1)}
\ifnum\lastpenalty>9000\else\bigbreak\fi
\noindent{\it\secsym\the\subsecno. #1}\writetoca{\string\quad 
{\secsym\the\subsecno.} {#1}}\par\nobreak\medskip\nobreak}
\def\appendix#1#2{\global\meqno=1\global\subsecno=0\xdef\secsym{\hbox{#1.}}
\bigbreak\bigskip\noindent{\bf Appendix #1. #2}\message{(#1. #2)}
\writetoca{Appendix {#1.} {#2}}\par\nobreak\medskip\nobreak}
%
%
\def\eqnn#1{\xdef #1{(\secsym\the\meqno)}\writedef{#1\leftbracket#1}%
\global\advance\meqno by1\wrlabeL#1}
\def\eqna#1{\xdef #1##1{\hbox{$(\secsym\the\meqno##1)$}}
\writedef{#1\numbersign1\leftbracket#1{\numbersign1}}%
\global\advance\meqno by1\wrlabeL{#1$\{\}$}}
\def\eqn#1#2{\xdef #1{(\secsym\the\meqno)}\writedef{#1\leftbracket#1}%
\global\advance\meqno by1$$#2\eqno#1\eqlabeL#1$$}
%
\newskip\footskip\footskip14pt plus 1pt minus 1pt 
\def\footnotefont{\ninepoint}\def\f@t#1{\footnotefont #1\@foot}
\def\f@@t{\baselineskip\footskip\bgroup\footnotefont\aftergroup\@foot\let\next}
\setbox\strutbox=\hbox{\vrule height9.5pt depth4.5pt width0pt}
\global\newcount\ftno \global\ftno=0
\def\foot{\global\advance\ftno by1\footnote{$^{\the\ftno}$}}
%
\newwrite\ftfile   
\def\footend{\def\foot{\global\advance\ftno by1\chardef\wfile=\ftfile
$^{\the\ftno}$\ifnum\ftno=1\immediate\openout\ftfile=foots.tmp\fi%
\immediate\write\ftfile{\noexpand\smallskip%
\noexpand\item{f\the\ftno:\ }\pctsign}\findarg}%
\def\footatend{\vfill\eject\immediate\closeout\ftfile{\parindent=20pt
\centerline{\bf Footnotes}\nobreak\bigskip\input foots.tmp }}}
\def\footatend{}
%
%
\global\newcount\refno \global\refno=1
\newwrite\rfile
\def\ref{[\the\refno]\nref}
\def\nref#1{\xdef#1{[\the\refno]}\writedef{#1\leftbracket#1}%
\ifnum\refno=1\immediate\openout\rfile=refs.tmp\fi
\global\advance\refno by1\chardef\wfile=\rfile\immediate
\write\rfile{\noexpand\item{#1\ }\reflabeL{#1\hskip.31in}\pctsign}\findarg}
\def\findarg#1#{\begingroup\obeylines\newlinechar=`\^^M\pass@rg}
{\obeylines\gdef\pass@rg#1{\writ@line\relax #1^^M\hbox{}^^M}%
\gdef\writ@line#1^^M{\expandafter\toks0\expandafter{\striprel@x #1}%
\edef\next{\the\toks0}\ifx\next\em@rk\let\next=\endgroup\else\ifx\next\empty%
\else\immediate\write\wfile{\the\toks0}\fi\let\next=\writ@line\fi\next\relax}}
\def\striprel@x#1{} \def\em@rk{\hbox{}} 
\def\lref{\begingroup\obeylines\lr@f}
\def\lr@f#1#2{\gdef#1{\ref#1{#2}}\endgroup\unskip}
\def\semi{;\hfil\break}
\def\addref#1{\immediate\write\rfile{\noexpand\item{}#1}} 
\def\startrefs#1{\immediate\openout\rfile=refs.tmp\refno=#1}
\def\xref{\expandafter\xr@f}\def\xr@f[#1]{#1}
\def\refs#1{\count255=1[\r@fs #1{\hbox{}}]}
\def\r@fs#1{\ifx\und@fined#1\message{reflabel \string#1 is undefined.}%
\nref#1{need to supply reference \string#1.}\fi%
\vphantom{\hphantom{#1}}\edef\next{#1}\ifx\next\em@rk\def\next{}%
\else\ifx\next#1\ifodd\count255\relax\xref#1\count255=0\fi%
\else#1\count255=1\fi\let\next=\r@fs\fi\next}
%

%
\newwrite\ffile\global\newcount\figno \global\figno=1
\def\fig{fig.~\the\figno\nfig}
\def\nfig#1{\xdef#1{fig.~\the\figno}%
\writedef{#1\leftbracket fig.\noexpand~\the\figno}%
\ifnum\figno=1\immediate\openout\ffile=figs.tmp\fi\chardef\wfile=\ffile%
\immediate\write\ffile{\noexpand\medskip\noexpand\item{Fig.\ \the\figno. }
\reflabeL{#1\hskip.55in}\pctsign}\global\advance\figno by1\findarg}
\def\xfig{\expandafter\xf@g}\def\xf@g fig.\penalty\@M\ {}
\def\figs#1{figs.~\f@gs #1{\hbox{}}}
\def\f@gs#1{\edef\next{#1}\ifx\next\em@rk\def\next{}\else
\ifx\next#1\xfig #1\else#1\fi\let\next=\f@gs\fi\next}
\newwrite\lfile
{\escapechar-1\xdef\pctsign{\string\%}\xdef\leftbracket{\string\{}
\xdef\rightbracket{\string\}}\xdef\numbersign{\string\#}}

\def\writestop{\def\writestoppt{\immediate\write\lfile{\string\pageno%
\the\pageno\string\startrefs\leftbracket\the\refno\rightbracket%
\string\def\string\secsym\leftbracket\secsym\rightbracket%
\string\secno\the\secno\string\meqno\the\meqno}\immediate\closeout\lfile}}
\def\writestoppt{}\def\writedef#1{}
\def\seclab#1{\xdef #1{\the\secno}\writedef{#1\leftbracket#1}\wrlabeL{#1=#1}}
\def\subseclab#1{\xdef #1{\secsym\the\subsecno}%
\writedef{#1\leftbracket#1}\wrlabeL{#1=#1}}
\newwrite\tfile \def\writetoca#1{}
\def\leaderfill{\leaders\hbox to 1em{\hss.\hss}\hfill}
\def\writetoc{\immediate\openout\tfile=toc.tmp 
   \def\writetoca##1{{\edef\next{\write\tfile{\noindent ##1 
   \string\leaderfill {\noexpand\number\pageno} \par}}\next}}}

%
\catcode`\@=12 
%
\edef\tfontsize{\ifx\answ\bigans scaled\magstep3\else scaled\magstep4\fi}
\font\titlerm=cmr10 \tfontsize \font\titlerms=cmr7 \tfontsize
\font\titlermss=cmr5 \tfontsize \font\titlei=cmmi10 \tfontsize
\font\titleis=cmmi7 \tfontsize \font\titleiss=cmmi5 \tfontsize
\font\titlesy=cmsy10 \tfontsize \font\titlesys=cmsy7 \tfontsize
\font\titlesyss=cmsy5 \tfontsize \font\titleit=cmti10 \tfontsize
\skewchar\titlei='177 \skewchar\titleis='177 \skewchar\titleiss='177
\skewchar\titlesy='60 \skewchar\titlesys='60 \skewchar\titlesyss='60
\def\titlefont{\def\rm{\fam0\titlerm}
\textfont0=\titlerm \scriptfont0=\titlerms \scriptscriptfont0=\titlermss
\textfont1=\titlei \scriptfont1=\titleis \scriptscriptfont1=\titleiss
\textfont2=\titlesy \scriptfont2=\titlesys \scriptscriptfont2=\titlesyss
\textfont\itfam=\titleit \def\it{\fam\itfam\titleit}\rm}
 \ifx\answ\bigans\else scaled\magstep1\fi
\ifx\answ\bigans\def\abstractfont{\tenpoint}\else
\font\abssl=cmsl10 scaled \magstep1
\font\absrm=cmr10 scaled\magstep1 \font\absrms=cmr7 scaled\magstep1
\font\absrmss=cmr5 scaled\magstep1 \font\absi=cmmi10 scaled\magstep1
\font\absis=cmmi7 scaled\magstep1 \font\absiss=cmmi5 scaled\magstep1
\font\abssy=cmsy10 scaled\magstep1 \font\abssys=cmsy7 scaled\magstep1
\font\abssyss=cmsy5 scaled\magstep1 \font\absbf=cmbx10 scaled\magstep1
\skewchar\absi='177 \skewchar\absis='177 \skewchar\absiss='177
\skewchar\abssy='60 \skewchar\abssys='60 \skewchar\abssyss='60
\def\abstractfont{\def\rm{\fam0\absrm}
\textfont0=\absrm \scriptfont0=\absrms \scriptscriptfont0=\absrmss
\textfont1=\absi \scriptfont1=\absis \scriptscriptfont1=\absiss
\textfont2=\abssy \scriptfont2=\abssys \scriptscriptfont2=\abssyss
\textfont\itfam=\bigit \def\it{\fam\itfam\bigit}\def\footnotefont{\tenpoint}%
\textfont\slfam=\abssl \def\sl{\fam\slfam\abssl}%
\textfont\bffam=\absbf \def\bf{\fam\bffam\absbf}\rm}\fi
\def\tenpoint{\def\rm{\fam0\tenrm}
\textfont0=\tenrm \scriptfont0=\sevenrm \scriptscriptfont0=\fiverm
\textfont1=\teni  \scriptfont1=\seveni  \scriptscriptfont1=\fivei
\textfont2=\tensy \scriptfont2=\sevensy \scriptscriptfont2=\fivesy
\textfont\itfam=\tenit \def\it{\fam\itfam\tenit}\def\footnotefont{\ninepoint}%
\textfont\bffam=\tenbf \def\bf{\fam\bffam\tenbf}\def\sl{\fam\slfam\tensl}\rm}
\font\ninerm=cmr9 \font\sixrm=cmr6 \font\ninei=cmmi9 \font\sixi=cmmi6 
\font\ninesy=cmsy9 \font\sixsy=cmsy6 \font\ninebf=cmbx9 
\font\nineit=cmti9 \font\ninesl=cmsl9 \skewchar\ninei='177
\skewchar\sixi='177 \skewchar\ninesy='60 \skewchar\sixsy='60 
\def\ninepoint{\def\rm{\fam0\ninerm}
\textfont0=\ninerm \scriptfont0=\sixrm \scriptscriptfont0=\fiverm
\textfont1=\ninei \scriptfont1=\sixi \scriptscriptfont1=\fivei
\textfont2=\ninesy \scriptfont2=\sixsy \scriptscriptfont2=\fivesy
\textfont\itfam=\ninei \def\it{\fam\itfam\nineit}\def\sl{\fam\slfam\ninesl}%
\textfont\bffam=\ninebf \def\bf{\fam\bffam\ninebf}\rm} 
%
%

\hyphenation{anom-aly anom-alies coun-ter-term coun-ter-terms}
\def\inv{^{\raise.15ex\hbox{${\scriptscriptstyle -}$}\kern-.05em 1}}

\def\Dsl{\,\raise.15ex\hbox{/}\mkern-13.5mu D} 
\def\dsl{\raise.15ex\hbox{/}\kern-.57em\partial}

\font\bigit=cmti10 scaled \magstep1
\def\lspace{\ifx\answ\bigans{}\else\qquad\fi}
\def\lbspace{\ifx\answ\bigans{}\else\hskip-.2in\fi} 
\def\boxeqn#1{\vcenter{\vbox{\hrule\hbox{\vrule\kern3pt\vbox{\kern3pt
	\hbox{${\displaystyle #1}$}\kern3pt}\kern3pt\vrule}\hrule}}}
\def\mbox#1#2{\vcenter{\hrule \hbox{\vrule height#2in
		\kern#1in \vrule} \hrule}}  
%

\def\e#1{{\rm e}^{^{\textstyle#1}}}

\def\darr#1{\raise1.5ex\hbox{$\leftrightarrow$}\mkern-16.5mu #1}

\def\half{{\textstyle{1\over2}}} 
\def\roughly#1{\raise.3ex\hbox{$#1$\kern-.75em\lower1ex\hbox{$\sim$}}}

\lref\Aisaka{
  Y.~Aisaka and Y.~Kazama,
  ``Origin of pure spinor superstring,''
JHEP {\bf 0505}, 046 (2005).
[hep-th/0502208].
}

\lref\BerkovitsGH{
  N.~Berkovits,
  ``Pure spinors, twistors, and emergent supersymmetry,''
JHEP {\bf 1212}, 006 (2012).
[arXiv:1105.1147 [hep-th]].
}

\lref\BerkovitsC{
N.~Berkovits,
``Covariant Map Between Ramond-Neveu-Schwarz and Pure Spinor Formalisms for the Superstring,''
JHEP {\bf 04}, 024 (2014).
[arXiv:1312.0845 [hep-th]].
}

\lref\Matone{
  M.~Matone, L.~Mazzucato, I.~Oda, D.~Sorokin and M.~Tonin,
  ``The Superembedding origin of the Berkovits pure spinor covariant quantization of superstrings,''
Nucl.\ Phys.\ B {\bf 639}, 182 (2002).
[hep-th/0206104].
}

\lref\Bnon{
  N.~Berkovits,
  ``Pure spinor formalism as an N=2 topological string,''
JHEP {\bf 0510}, 089 (2005).
[hep-th/0509120].
}

\lref\BerkovitsPLA{
  N.~Berkovits,
  ``Dynamical twisting and the b ghost in the pure spinor formalism,''
JHEP {\bf 1306}, 091 (2013).
[arXiv:1305.0693 [hep-th]].
}

\lref\Bpure{
  N.~Berkovits,
  ``Super Poincare covariant quantization of the superstring,''
JHEP {\bf 0004}, 018 (2000).
[hep-th/0001035].
}

\lref\Donagi{
  R.~Donagi and E.~Witten,
``Supermoduli Space Is Not Projected,''
[arXiv:1304.7798 [hep-th]]\semi
  E.~Witten,
  ``More On Superstring Perturbation Theory,''
[arXiv:1304.2832 [hep-th]]\semi
  E.~Witten,
  ``Superstring Perturbation Theory Revisited,''
[arXiv:1209.5461 [hep-th]].
}

\lref\explaining {
N.~Berkovits,
``Explaining the Pure Spinor Formalism for the Superstring,''
JHEP {\bf 0801}, 065 (2008).
[arXiv:0712.0324 [hep-th]].
}

\lref\FriedanGE{
  D.~Friedan, E.~J.~Martinec and S.~H.~Shenker,
  ``Conformal Invariance, Supersymmetry and String Theory,''
Nucl.\ Phys.\ B {\bf 271}, 93 (1986).
}

\lref\Baulieua{
 L.~Baulieu,
 ``Transmutation of pure 2-D supergravity into topological 2-D gravity and other conformal theories,''
Phys.\ Lett.\ B {\bf 288}, 59 (1992).
[hep-th/9206019]\semi
  L.~Baulieu, M.~B.~Green and E.~Rabinovici,
  ``A Unifying topological action for heterotic and type II superstring theories,''
Phys.\ Lett.\ B {\bf 386}, 91 (1996).
[hep-th/9606080]\semi
 L.~Baulieu, M.~B.~Green and E.~Rabinovici,
  ``Superstrings from theories with $N>1$ world sheet supersymmetry,''
Nucl.\ Phys.\ B {\bf 498}, 119 (1997).
[hep-th/9611136]\semi
  L.~Baulieu and N.~Ohta,
  ``World sheets with extended supersymmetry,''
Phys.\ Lett.\ B {\bf 391}, 295 (1997).
[hep-th/9609207].
}

\lref\osv{
  O.~Chandia,
  ``The b Ghost of the Pure Spinor Formalism is Nilpotent,''
Phys.\ Lett.\ B {\bf 695}, 312 (2011).
[arXiv:1008.1778 [hep-th]].
}

\lref\WittenMB{
  E.~Witten,
  ``D = 10 Superstring Theory,''
In *Philadelphia 1983, Proceedings, Grand Unification*, 395-408.
}

\lref\rennan{
  R.~L.~Jusinskas,
  ``Nilpotency of the b ghost in the non minimal pure spinor formalism,''
[arXiv:1303.3966 [hep-th]].
}

\lref\BeisertJR{
  N.~Beisert, C.~Ahn, L.~F.~Alday, Z.~Bajnok, J.~M.~Drummond, L.~Freyhult, N.~Gromov and R.~A.~Janik {\it et al.},
  ``Review of AdS/CFT Integrability: An Overview,''
Lett.\ Math.\ Phys.\  {\bf 99}, 3 (2012).
[arXiv:1012.3982 [hep-th]].
}

\lref\Grassia{
  P.~A.~Grassi, G.~Policastro, M.~Porrati and P.~Van Nieuwenhuizen,
  ``Covariant quantization of superstrings without pure spinor constraints,''
JHEP {\bf 0210}, 054 (2002).
[hep-th/0112162]\semi
  P.~A.~Grassi, G.~Policastro and P.~van Nieuwenhuizen,
  ``On the BRST cohomology of superstrings with / without pure spinors,''
Adv.\ Theor.\ Math.\ Phys.\  {\bf 7}, 499 (2003).
[hep-th/0206216].
}

\lref\Matone{
  M.~Matone, L.~Mazzucato, I.~Oda, D.~Sorokin and M.~Tonin,
  ``The Superembedding origin of the Berkovits pure spinor covariant quantization of superstrings,''
Nucl.\ Phys.\ B {\bf 639}, 182 (2002).
[hep-th/0206104].
}

\lref\Tonin{
  M.~Tonin,
  ``World sheet supersymmetric formulations of Green-Schwarz superstrings,''
Phys.\ Lett.\ B {\bf 266}, 312 (1991).
}

\lref\Sorokin{
  D.~P.~Sorokin,
  ``Superbranes and superembeddings,''
Phys.\ Rept.\  {\bf 329}, 1 (2000).
[hep-th/9906142].
}

\lref\Aisakaa{
  Y.~Aisaka and Y.~Kazama,
  ``A new first class algebra, homological perturbation and extension of pure spinor formalism for superstring,''
JHEP {\bf 0302}, 017 (2003).
[hep-th/0212316]\semi
  Y.~Aisaka and Y.~Kazama,
  ``Operator mapping between RNS and extended pure spinor formalisms for superstring,''
JHEP {\bf 0308}, 047 (2003).
[hep-th/0305221].
}

\lref\BerkovitsWR{
  N.~Berkovits,
  ``Covariant quantization of the Green-Schwarz superstring in a Calabi-Yau background,''
Nucl.\ Phys.\ B {\bf 431}, 258 (1994).
[hep-th/9404162].
}

\lref\BerkovitsUS{
  N.~Berkovits,
  ``Relating the RNS and pure spinor formalisms for the superstring,''
JHEP {\bf 0108}, 026 (2001).
[hep-th/0104247].
}

\lref\kroyter{
M.~Kroyter,
``Superstring field theory in the democratic picture,''
Adv. Theor. Math. Phys. {\bf 15}, no.3, 741-781 (2011).
[arXiv:0911.2962 [hep-th]].}

\lref\BerkovitsU{
  N.~Berkovits,
  ``Quantization of the superstring with manifest U(5) superPoincare invariance,''
Phys.\ Lett.\ B {\bf 457}, 94 (1999).
[hep-th/9902099].
}

\lref\BerkovitsVI{
  N.~Berkovits and N.~Nekrasov,
  ``Multiloop superstring amplitudes from non-minimal pure spinor formalism,''
JHEP {\bf 0612}, 029 (2006).
[hep-th/0609012].
}

\lref\SiegelT{
  W.~Siegel,
  ``Superfields in Higher Dimensional Space-time,''
Phys.\ Lett.\ B {\bf 80}, 220 (1979).
}

\lref\WittenT{
  E.~Witten,
  ``Twistor - Like Transform in Ten-Dimensions,''
Nucl.\ Phys.\ B {\bf 266}, 245 (1986).
}

\lref\BerkovitsBT{
  N.~Berkovits,
  ``Pure spinor formalism as an N=2 topological string,''
JHEP {\bf 0510}, 089 (2005).
[hep-th/0509120].
}

\lref\BerkovitsNek{
  N.~Berkovits and N.~Nekrasov,
  ``Multiloop superstring amplitudes from non-minimal pure spinor formalism,''
JHEP {\bf 0612}, 029 (2006).
[hep-th/0609012].
}

\lref\siegelclassical{
  W.~Siegel,
  ``Classical Superstring Mechanics,''
Nucl.\ Phys.\ B {\bf 263}, 93 (1986).
}

\lref\Baulieuf{
  L.~Baulieu,
  ``SU(5)-invariant decomposition of ten-dimensional Yang-Mills supersymmetry,''
Phys.\ Lett.\ B {\bf 698}, 63 (2011).
[arXiv:1009.3893 [hep-th]].
}

\lref\SiegelYD{
  W.~Siegel,
[arXiv:1005.2317 [hep-th]].
}

\def\bar{\overline}

\def\a{{\alpha}}

\def\l{{\lambda}}
\def\lb{{\overline\lambda}}
\def\wb{{\overline w}}
\def\wtb{{\widetilde{\overline w}}}

\def\whb{\widehat{\overline w}}

\def\lb{{\overline\lambda}}
\def\b{{\beta}}
\def\bh{{\widehat\beta}}

\def\g{{\gamma}}
\def\G{{\Gamma}}
\def\Gb{{\bar\Gamma}}
\def\gh{{\widehat\gamma}}
\def\etah{{\widehat\eta}}
\def\phih{{\widehat\phi}}
\def\xih{{\widehat\xi}}

\def\d{{\delta}}
\def\e{{\epsilon}}

\def\L{{\Lambda}}
\def\O{{\Omega}}
\def\half{{1\over 2}}
\def\p{{\partial}}

\def\pb{{\overline\partial}}
\def\t{{\theta}}

\def\S{{\Sigma}}

\def\lb{{\bar{\lambda}}}

\Title{\vbox{\baselineskip12pt
\hbox{}}}
{{\vbox{\centerline{Manifest Spacetime Supersymmetry}
\smallskip
\centerline{and the Superstring}}} }
\bigskip\centerline{Nathan Berkovits\foot{e-mail: nathan.berkovits@unesp.br}}
\bigskip
\centerline{\it ICTP South American Institute for Fundamental Research}
\centerline{\it Instituto de F\'\i sica Te\'orica, UNESP - Univ. 
Estadual Paulista }
\centerline{\it Rua Dr. Bento T. Ferraz 271, 01140-070, S\~ao Paulo, SP, Brasil}
\bigskip

\vskip .3in

The algebra of spacetime supersymmetry generators in the RNS formalism for the superstring closes only up to a picture-changing operation. After adding non-minimal variables and working in the ``large" Hilbert space, the algebra closes without picture-changing and spacetime supersymmetry can be made manifest. The resulting non-minimal version of the RNS formalism is related by a field redefinition to the pure spinor formalism.

\vskip .3in

\Date {June 2021}
\newsec{Introduction}

Although the RNS formalism for the superstring has a beautiful geometric foundation coming from worldsheet super-reparameterization invariance, the lack of manifest spacetime supersymmetry
complicates the computation of multiloop scattering amplitudes and the description of Ramond-Ramond backgrounds. As shown in \FriedanGE, spacetime supersymmetry generators $q_\alpha$ can be constructed in this formalism which satisfy the algebra $\{q_\a, q_\b\} = \g^m_{\a\b} \tilde P_m $ where $\tilde P_m = \int dz ~e^{-\phi}\psi_m$ is in the $-1$ picture. But since the picture-changing operation which relates $\tilde P_m$ with the usual
translation generator $P_m = \int dz~\p x_m$ is only valid on onshell states, this algebra only closes onshell and spacetime supersymmetry is not manifest.

On the other hand, the pure spinor formalism for the superstring \Bpure\ has manifest spacetime supersymmetry and can describe Ramond-Ramond backgrounds, but lacks a geometric foundation where the pure spinor BRST operator comes from gauge-fixing a worldsheet symmetry. There is complete agreement between all scattering amplitudes which have been computed using the two formalisms, but equivalence has not yet been proven and multiloop amplitudes in the two formalisms have different types of subtleties.

In this paper, the RNS and pure spinor formalisms will be shown to be equivalent by relating them to a manifestly spacetime supersymmetric generalization of the RNS formalism. In addition to the usual RNS matter and ghost variables, this generalization will involve non-minimal variables that include both pure spinor variables and the fermionic $\theta^\a$ and conjugate momenta $d_\alpha$ variables of the Green-Schwarz-Siegel formalism.  Since the new formalism includes features of the pure spinor, RNS and Green-Schwarz-Siegel formalisms, it will be called the B-RNS-GSS formalism. 

The first step to making spacetime supersymmetry manifest in the RNS formalism is to work in the ``large" Hilbert space including the ghost zero modes $(\xi_0, \eta_0)$ and to define the extended BRST operator $Q'  = Q_{RNS} - \eta_0$. As discussed in \BerkovitsUS\ and \kroyter, picture-changing can be interpreted as a BRST-trivial gauge transformation using $Q'$ in the large Hilbert space. 
The next step is to add non-minimal variables to the RNS formalism and construct an operator $\rho_\a$ such that $\{q'_\a, q'_\b\} = \g^m_{\a\b} \int dz~\p x_m$ where 
$q'_\a = q^{RNS}_\a + Q'(\rho_\a)$. And the final step is to find a similarity transformation which maps $q'_\a$ into the spacetime supersymmetry generators of the Green-Schwarz-Siegel formalism. 

The resulting worldsheet action and BRST operator for this B-RNS-GSS formalism are manifestly spacetime supersymmetric, and BRST-invariant vertex operators are easily constructed in terms of d=10 superfields. Furthermore,
a field redefinition is found which maps this worldsheet action, BRST operator and vertex operators into the worldsheet action, BRST operator and vertex operators of the pure spinor formalism. 
It is hoped that by studying this intermediate B-RNS-GSS formalism, the relation between the multiloop subtleties in the RNS and pure spinor formalisms will be better understood.

\newsec{B-RNS-GSS Formalism}

The B-RNS-GSS worldsheet action will be constructed from the usual RNS matter and ghost variables
$(x^m, \psi^m; b, c, \b, \g)$ for $m=0$ to 9, together with the Green-Schwarz-Siegel fermionic spacetime spinor variables $(\t^\a, p_\a)$ for $\a=1$ to 16, the unconstrained bosonic spacetime spinor variables $(\L^\a, \O_\a)$, and the bosonic and fermionic pure spinor variables $(\lb_\a, \wb^\a)$ and $(r_\a, s^\a)$ satisfying the pure spinor constraints $\lb\g^m\lb = \lb \g^m r =0$.

The left-moving contribution to the worldsheet action  and stress tensor are 
\eqn\action{S_{B-RNS-GSS} = S_{RNS} + \int d^2 z [ \O_\a \pb \L^\a + \wb_\a \pb \lb^\a + p_\a \pb \t^\a + s^\a \pb r_\a ],}
\eqn\stress{T_{B-RNS-GSS} = T_{RNS} - \O_\a \p\L^\a - \wb_\a \p\lb^\a - p_\a \p\t^\a - s^\a \p r_\a,} 
where the RNS worldsheet action, stress tensor, and BRST operator are defined as
\eqn\rnsa{S_{RNS} = \int d^2 z (\half \p x^m \bar\p x_m +\half \psi^m \bar\p \psi_m
+\b\bar\p \g + b\bar\p c),}
\eqn\rnsb{ T_{RNS} = -\half \p x^m \p x_m - \half\psi^m \p \psi_m
-\b\p \g -\half \p(\b\g) - b\p c - \p(b c),}
\eqn\rnsc{Q_{RNS} = \int dz ( c T_{RNS} + \g \psi^m \p x_m + \g^2 b - b c \p c ),} and the right-moving variables
will be ignored throughout this paper.
Although only the open superstring will be discussed
in this paper, all results can be easily generalized
to the closed superstring 
by taking the ``left-right
product'' of two open superstrings.

The natural BRST operator for this extended formalism is 
\eqn\brstone{Q_{nonmin} = Q_{RNS} + \int dz [ \L^\a p_\a + \wb^\a r_\a]}
since the non-minimal term $\int dz [ \L^\a p_\a + \wb^\a r_\a]$ implies that states in the cohomology of
$Q_{nonmin}$ are states in the cohomology of $Q_{RNS}$ which are independent of the spacetime
spinor variables $[\L^\a, \O_\a, \t^\a, p_\a, \lb_\a, \wb^\a, r_\a, s^\a]$.

In the RNS formalism \FriedanGE, the spacetime supersymmetry generator in the $-\half$ picture is
$q_\a = \int dz ˜e^{-\half \phi} \S_\a$ where $\S_\a$ is the spin field of conformal weight $5\over 8$ constructed from $\psi^m$ and the $(\b, \g)$ ghosts have been fermionized as 
$\b = \p \xi e^{-\phi}$ and $\g = \eta e^\phi$. Since $\{q_\a, q_\b\} = \g^m_{\a\b} \int dz ~e^{-\phi} \psi_m$,
the spacetime supersymmetry algebra only closes up to picture-changing. Note that $\{Q_{RNS}, \int dz
~\xi e^{-\phi} \psi^m\} = \int dz ~ \p x^m$, so $\int dz ~e^{-\phi} \psi^m$ is the translation generator in the $-1$
picture.

As discussed in \BerkovitsUS\ and \kroyter, picture-changing can be treated as a BRST-trivial operation if one extends the space
of states to the ``large" Hilbert space which includes the zero mode of $\xi$ and defines the BRST operator
as 
\eqn\newbrst{Q' = Q_{RNS} -\eta_0}
where $\eta_0 = \int dz~\eta$.
Picture-changing from $V$ to
$\{Q_{RNS}, \xi V\}$ is expressed by the BRST-trivial operation $V \to V + Q'(\xi V)= Q_{RNS} (\xi V)$. And it is easy to show that any state in the cohomology of $Q'$ can be expressed
as a state in the small Hilbert space in the cohomology of $Q_{RNS}$. 
For example, suppose that $Q'V=0$ where $V = V_P + V_{P+1}$ and $V_P$ carries picture $P$. Then $Q'V=0$ implies that 
\eqn\ver{\eta_0 V_P = Q_{RNS} V_{P} - \eta_0 V_{P+1} = Q_{RNS} V_{P+1} =0.}
So $V'= V + Q' (\xi V_P)$ satisfies $Q_{RNS} V'= \eta_0 V'=0$.

By working in the large Hilbert space with $Q' = Q_{nonmin} - \eta_0$ and shifting $q_\a$ by a BRST-trivial quantity, one can now make spacetime supersymmetry manifest by defining the generator
\eqn\newsusy{ q_\a = \int dz[ e^{-\half \phi} \S_\a + Q' (\rho_\a) ]}
where $\rho_\a$ is chosen such that $\{q_\a, q_\b\} = \int dz ~\p x_m \g^m_{\a\b}$. Furthermore, after performing a similarity transformation $q_\a \to e^{-R} q_\a e^{R}$, this generator can be mapped to the
Green-Schwarz-Siegel supersymmetry generator\siegelclassical\ 
\eqn\susyg{q^{GSS}_\a = \int dz [ p_\a + \half (\p x^m +{1\over{12}}\t\g^m \p\t)(\g_m\t)_\a ].} 

The resulting BRST operator $Q_{B-RNS-GSS} = e^{-R} ( Q_{nonmin} - \eta_0) e^R$ is manifestly spacetime supersymmetric and is 
\eqn\brstfinal{Q_{B-RNS-GSS} = 
\int dz [\L^\a d_\a +\wb^\a r_\a +\half \xi e^{-\phi} (\L\g^m\L)\psi_m 
- e^{-\half \phi} \L^\a\S_\a} 
$$+  \g \psi^m \Pi_m + \g^2 (b+\O_\a\p\t^\a + s^\a \p\lb_\a) + c T - b c \p c ] - \eta_0$$
where 
\eqn\defd{d_\a = p_\a -\half (\p x^m +{1\over 4} \t\g^m \p\t)(\g_m\t)_\a , \quad
\Pi^m = \p x^m +\half\t\g^m \p\t}
are the usual GSS spacetime supersymmetric operators \siegelclassical\ for fermionic and bosonic momenta.
Note that $T$ of \stress\ can be written in the manifestly spacetime supersymmetric form 
\eqn\tsusy{T = -\half \Pi^m \Pi_m - d_\a \p\t^\a -  \O_\a \p\L^\a -\wb^\a \p\lb_\a - s^\a \p r_\a - \half\psi^m \p \psi_m
-\b\p \g -\half \p(\b\g) - b\p c - \p(b c).}

To find the shift $\rho_\a$ and similarity transformation $R$ that map the supersymmetry generator and BRST operator into \susyg\ and \brstfinal, first consider the shift and similarity transformation
\eqn\shift{ \tilde\rho_\a = -\O_\a +\half \xi e^{-\phi} \psi^m (\g_m \t)_\a, }
\eqn\similarityone{ \tilde R = \int dz [ c (\O_\a \p\t^\a + s^\a\p\lb_\a)+ e^{-\half\phi} \S_\a \t^\a+ \half\xi e^{-\phi}
(\t\g^m \L) \psi_m].}
One finds that up to terms proportional to $\t^\a$,  $ e^{-\tilde R} Q' e^{\tilde R} $ is equal to $Q_{B-RNS-GSS}$ of \brstfinal\ and 
$e^{-\tilde R} \int dz[ e^{-\half \phi} \S_\a + Q' (\tilde\rho_\a) ] e^{\tilde R}$ is equal to $q^{GSS}_\a$ of \susyg\ where
$Q' = Q_{nonmin} - \eta_0$. But since $\t^\a$ is a non-minimal variable, cohomology arguments imply that these terms proportional to $\t^\a$ can be written as $[Q', f (\theta)]$ for some function $f(\theta)$ which is at least quadratic order in $\t^\a$.
Therefore, $Q_{B-RNS-GSS} = e^{-R} Q' e^R$ and 
$q^{GSS}_\a = e^{- R} \int dz[ e^{-\half \phi} \S_\a + Q' (\rho_\a) ] e^{R}$
where
$R = \tilde R  + f(\theta)$, $\rho_\a = \tilde \rho_\a + f_\a(\theta)$,  and $f$ and $f_\alpha$ are terms of quadratic or higher-order in $\t^\a$.  

Using the similarity transformation $R$, vertex operators in the cohomology of $Q_{B-RNS-GSS}$ can be obtained from the usual RNS vertex operators as $V= e^{-R} V_{RNS} e^R$. However, it is simpler to solve the condition that $Q_{B-RNS-GSS} V =0$ and one finds that the unintegrated and integrated massless vertex operators are \BerkovitsC
\eqn\app{V = \L^\a A_\a (x,\t) - \g\psi^m A_m(x,\t) -\g^2 \O_\a W^\a (x,\t)} 
$$+ c [\p\t^\a A_\a(x,\t) + \Pi^m A_m (x,\t) + d_\a W^\a (x,\t) $$
$$-\half (\psi^m\psi^n - \half \L \g^{mn}\O) F_{mn}(x,\t) -
\g\psi^m \O_\a \p_m W^\a(x,\t)] ,$$
\vskip 5pt
\eqn\intu{\int dz ~U = \int dz [ \p\t^\a A_\a(x,\t) + \Pi^m A_m (x,\t) + d_\a W^\a (x,\t)}
$$+\half
 (-\psi^m\psi^n + \half \L \g^{mn}\O) F_{mn}(x,\t) -
 \g\psi^m \O_\a \p_m W^\a(x,\t)],$$
where $(A_\a, A_m, W^\a, F_{mn})$ are the usual $d=10$ super-Yang-Mills superfields satisfying
\eqn\onshell{D_\a A_\b + D_\b A_\a = \g^m_{\a\b} A_m, \quad
D_\a A_m - \p_m A_\a = (\g_m)_{\a\b} W^\b, }
$$D_\a W^\b = \half (\g^{mn})_\a{}^\b \p_m A_n = {1\over 4}(\g^{mn})_\a{}^\b F_{mn}, $$
and  $D_\a = {\p\over{\p\t^\a}} + \half (\g^m\t)_\a \p_m$ is the $d=10$ supersymmetric
derivative. 

Note that the similarity transformation of \similarityone\ mixes different pictures, but one can choose a representative of the BRST cohomology so that the massless super-Yang-Mills vertex operator $V$ of \app\ is in the zero picture. And by constructing massive superstring vertex operators from the OPE's of massless vertex operators, one can similarly find vertex operators at zero picture for all GSO-projected
onshell superstring states. For states which are not GSO-projected, i.e. which have square-root cuts
with $e^{-\half \phi} \S_\a$, the vertex operator will need to involve spin fields in $\L^\a$ and $\t^\a$ so that the operator does not have square-root cuts with the term $\int dz~\L^\a ( d_\a - e^{-\half \phi} \S_\a)$ in the BRST operator of \brstfinal.

For the gluon component of the superfield with polarization $\e_m$, one finds from \app\ that
\eqn\vertexgluon{V = \e_m e^{i k\cdot x} [V^m_{RNS} - {i\over 2} c k_n (\L \g^{mn} \O + \t \g^{mn} p) + {\cal O}^m(\t)] }
where $V^m_{RNS} = -\g\psi^m + c (\p x^m -i k_n \psi^n \psi^m)$ is the usual RNS vertex operator
and ${\cal O}^m(\t)$ denotes terms which are linear or higher-order in $\t^\a$. It is easy to verify
that the terms ${\cal O}^m(\t)$ and $(\L \g^{mn} \O + \t \g^{mn} p)$ in \vertexgluon\ decouple from gluon 
scattering amplitudes since there are no extra $d_\a$'s to contract with the ${\cal O}^m(\t)$ term
and since contractions of $(\L \g^{mn} \O + \t \g^{mn} p)$ with itself cancel because of opposite
signs from the bosonic and fermionic contributions. So the scattering amplitudes for external gluons
in the B-RNS-GSS formalism are identical to the amplitudes in the usual RNS formalism.

However, the vertex operator for the gluino component of the superfields in \app\ is very different from
the gluino vertex operator in the RNS formalism. Just as the spacetime supersymmetry generator
of \susyg\ does not involve the spin field $\S_\a$, the gluino vertex operator in the B-RNS-GSS
formalism does not involve the spin field. So the amplitude computation for external fermions looks
very different from the RNS computation. Nevertheless, the two computations are guaranteed to give
equivalent results since they are related by a similarity transformation.

\newsec{Relation with Pure Spinor Formalism}

In this section, the BRST operator of \brstfinal\ in the B-RNS-GSS formalism will be mapped to
the pure spinor BRST operator by a field redefinition and similarity transformation. The field redefinition 
will ``dynamically twist" the RNS variables $\psi^m$ of spin $\half$ and $(\b,\g)$ of spin $({3\over 2}, -\half)$ into the variables $(\G^m, \Gb_m)$ of spin $(0,1)$ and $(\bh, \gh)$ of spin $(2,-1)$ \BerkovitsPLA. The twisting of the $\psi^m$ variables to $(\G^m, \Gb_m)$ variables
shifts their central charge contribution from
$+5$ to $-10$, and is compensated by the replacement of the $(\b,\g)$ with $(\bh,\gh)$ variables which shifts their central charge contribution from $+11$ to $+26$. After performing a similarity transformation, the BRST operator of \brstfinal\ in terms of these twisted variables is mapped to
$Q_{pure} - \etah_0$ where $Q_{pure} = \int dz (\l^\a d_\a + \wb^\a r_\a)$ is the pure spinor BRST operator and $\etah_0$ is the zero mode coming from fermionizing $\bh = \p\xih e^{-\phih}$ and
$\gh = \etah e^{\phih}$.

To dynamically twist the ten $\psi^m$ spin-half variables into five spin-zero and five
spin-one variables, it will be useful to construct a pure spinor $\l^\a$ out of $\L^\a$ and $\lb_\a$ as
\eqn\lam{\l^\a = \L^\a - {1\over{2(\L\lb)}} (\L\g^m\L) (\g_m\lb)^\a.} So $\l^\a$ and $\lb_\a$ satisfy
$\l\g^m \l = \lb\g^m\lb =0$, and their 11 complex components (in Wick-rotated Euclidean space) parameterize the complex space
${{SO(10)}\over {U(5)}}\times C$. Using the pure spinor variables $(\l^\a,\lb_\a)$ to covariantly choose the direction of the twisting, one can now dynamically twist the ten spin-half $\psi^m$ variables to spin-zero $\G^m$ variables and spin-one $\Gb^m$ variables defined by
\eqn\defgam{\G^m =  {1\over{2(\l\lb)} }\g (\l\g^m\g^n\lb) \psi_n, \quad
\Gb^m =  {1\over{2(\l\lb)}} {1\over \g} (\lb\g^m\g^n\l) \psi_n,}
so that 
\eqn\defpsi{\psi^m = \g \Gb^m + {1\over\g} {{(\l\g^m\g^n\lb)}\over{2(\l\lb)}} \G_n.}
$\bar\G^m$ will be constrained to satisfy $\Gb^m (\g_m\lb)^\a =0$, 
and
since $\psi^m$ of \defpsi\ is invariant under the gauge transformation $\d\G_m = \e\g_m\lb$
generated by this constraint, 
only half of the $\G^m$ and $\Gb^m$ components are
independent.

After expressing $\psi^m$ in terms of $\G^m$ and $\Gb^m$,
GSO-projected states only depend on even powers of the $\g$ ghost.
So it will be useful to define 
\eqn\ghdef{\gh \equiv (\g)^2}
which carries conformal weight $-1$, and define $\bh$ of conformal weight $+2$ to
be the conjugate momentum to $\gh$. Fermionizing $(\gh, \bh)$ as $\gh = \etah e^{\phih}$ and
$\bh = \p\xih e^{-\phih}$ and requiring that $(\etah, \xih, \phih)$ have no poles with $\G^m$ or $\Gb_m$, one finds that
\eqn\etahs{\etah = e^{-\half \phi}\l^\a \S_\a, \quad
\xih = e^{\half\phi} {1\over {(\l\lb)}} \lb_\a \S^\a, }
$$
e^{\phih} = \eta \p\eta e^{{5\over 2} \phi} {1\over {(\l\lb)}} \lb_\a \S^\a, \quad
e^{-\phih} = \xi \p\xi e^{-{5\over 2} \phi}  \l^\a \S_\a.$$
One can also express the unhatted ghost variables in terms of the twisted variables as
\eqn\etas{\eta = e^{-2\phih} {{(\lb \g_{mnpqr}\lb)}\over{120(\l\lb)^2}}\G^m \G^n\G^p \G^q \G^r, \quad
\xi = {1\over{120}}e^{2\phih} (\l \g^{mnpqr}\l)\Gb_m \Gb_n\Gb_p \Gb_q \Gb_r.}

In terms of these twisted variables, the BRST operator of \brstfinal\ is
\eqn\brstfinal{Q_{B-RNS-GSS} = 
\int dz [\l^\a d_\a +\wb^\a r_\a +\G^m {{(\lb\g_m\g_n\l)}\over{2(\l\lb)}}\Pi^n- \etah  - e^{-2\phih} {{(\lb \g_{mnpqr}\lb)}\over{120(\l\lb)^2}}\G^m \G^n\G^p \G^q \G^r }
$$ + u^m (\Gb_m + {{(\lb \g_m d)}\over {2(\l\lb)}} - \xih e^{-2\phih}{{ (\lb\g_{mnpqr}\lb)
}\over{24(\l\lb)^2}}\G^n\G^p\G^q\G^r)$$ 
$$+ \gh (b + \Gb_m \Pi^m + \O_\a\p\t^\a + s^\a \p\lb_\a) 
+ c T - b c \p c ] $$
where $u_m$ is defined by $\L^\a = \l^\a + {1\over{2(\l\lb)}} u^m (\g_m\lb)^\a$,
\eqn\stressthr{T = -\half \p x^m \p x_m -\Gb_m \p \G^m - \bh\p\gh - \p(\bh \gh) -
b\p c - \p (bc) }
$$ - p_\a \p\t^\a -  v_m \p u^m - w_\a \p \l^\a - \whb^{\a} \p \lb_\a - s^\a \p r_\a, $$
$\bh = {1\over{2\g}} \b - {1\over{2\g^2}} \Gb_m\G^m$,
$v_m$ and $\Gb_m$ are constrained to satisfy $v_m (\g^m \lb)^\a = \Gb_m (\g^m \lb)^\a =0$, and
$w_\a$ and $\whb^{\a}$ are defined to have no poles with each other or with $(\Gb_m, \G^m)$. In terms of
$w_\a$ and $\whb^{\a}$, one finds that
\eqn\defomega{\Omega_\a = {1\over{4(\l\lb)}}[\half (\lb\g^{mn})_\a (\l\g_{mn}w)+ \lb_\a (-\l^\b w_\b + 4 u_m v^m) ] }
$$+
{1\over {\gh}}{{(\l\lb)(\lb\g_m\g_n)_\a 
+ \lb_\a (\l\g_m\g_n\lb) } \over{4(\l\lb)^2}}\G^m \G^n + 
(\l\g^m)_\a v_m ,$$
\eqn\defwbar{\wb^\a = \whb^{\a} +{1\over{2(\l\lb)}}[ -u^m (\g_m \O)^\a  + 2 v_m u^m \l^\a - (\l\g_m\g_n)^\a
(\G^m \Gb^n - \half \gh  \Gb^m \Gb^n) ] .}

The next step to relating \brstfinal\ to the pure spinor BRST operator is to perform the similarity
transformation $Q_{B-RNS-GSS} \to  e^{-R'} Q_{B-RNS-GSS} e^{R'}$ where
\eqn\similtwo{R'  = \int dz ~[ -\xih e^{-2\phih} {{(\lb \g_{mnpqr}\lb)}\over{120(\l\lb)^2}}\G^m \G^n\G^p \G^q \G^r + \xih e^{-\phih}
{{(\lb \g_{mnp}r)}\over{24(\l\lb)^2}}\G^m \G^n\G^p ] 
.}
One finds that
\eqn\brstf{ e^{-R'} Q_{B-RNS-GSS} e^{R'} =}
$$
 \int dz [\l^\a d_\a +{\wtb}^{\a} r_\a + c T - b c \p c - \etah
+ \G^m {{(\lb\g_m\g_n\l)}\over{2(\l\lb)}}\Pi^n$$
$$
+\G^m \G^n [{{(\lb \g_{mnp}r)}\over{8(\l\lb)^2}}\Pi^p  + {{(r \g_{mnp}r)}\over{8(\l\lb)^2}}\Gb^p] 
+\G^m \G^n\G^p [ {{(\lb \g_{mnp}\p\lb)}\over{24(\l\lb)^2}}- (\lb\t){{(\lb \g_{mnp}r)}\over{12(\l\lb)^2}}]
 $$
 $$+ \gh (b+ \Gb^m \Pi_m + {{\Gb_m \Gb_n}\over{4(\l\lb)}} (\l\g^{mn} r) + s^\a \p\lb_\a +\O_\a\p\t^\a) $$
 $$+ u_m (\Gb^m +{1\over{2(\l\lb)}}\lb\g^m d - {{(\lb\g^{m} \g^{np} r)}\over{8(\l\lb)^2}} N_{np})] $$
 where 
 \eqn\defwtb{\wtb^{\a} \equiv \whb^{\a} + {1\over{2(\l\lb)}} (\l\g_m\g_n)^\a (u_m v_n - \G_m\Gb_n), \quad N_{mn} = 
\half (\l\g_{mn} w).}
 Since $\wtb^{\a}$ of \defwtb\  commutes with the constraints 
 $v^m (\g_m\lb)^\a = \Gb^m (\g_m\lb)^\a=0$ 
 up to the gauge transformation
 $\d \wtb^{\a} = f^m (\lb\g_m)^\a$, one can easily verify
 that \brstf\ also commutes with these constraints.
  
 The BRST operator of \brstf\ is constructed out of the operators  discussed in \BerkovitsPLA\ where the last line of \brstf\ is $u_m$ times
 the constraint in \BerkovitsPLA\ for $\Gb^m$ and the second-to-last line of \brstf\ is
 $\gh$ times the composite $b$ ghost expressed in terms of $\Gb^m$. 
After applying the similarity transformation 
${\cal O} \to e^{-R'''} e^{-R''} {\cal O} e^{R''} e^{R'''}$ where
\eqn\simthree{R'' = \int dz  {1\over{2(\l\lb)}}
\G^m [-\lb\g^m d + {{(\lb\g^{m} \g^{np} r)}\over{4(\l\lb)}} N_{np} ],}
$$ R'''=  \int dz \gh v^m (\Pi_m + {{(\l\g^m\g^n r)}\over{4(\l\lb)}}\Gb_n),$$
\brstf\ reduces to
\eqn\brstg{ e^{-R'''} e^{-R''} e^{-R'} Q_{B-RNS-GSS} e^{R'} e^{R''} e^{R'''} =} 
 $$\int dz (\l^\a d_\a +\wtb^{\a} r_\a - \etah+ \gh [b-B + v_m(\l\g^m)_\a \p ({{\G_n (\g^n\lb)^\a}\over{2(\l\lb)}} )] + u_m \Gb^m + c T - b c \p c)$$
 where $B$ is the usual composite pure spinor $b$ ghost (ignoring normal ordering terms)
 \eqn\compb{B = 
 - s^\a\p\lb_\a  + {{\lb_\a (2
\Pi^m (\g_m d)^\a-  N_{mn}(\g^{mn}\p\t)^\a
+\l^\b w_\b \p\t^\a)}\over{4(\lb\l)}} }
$$- {{(\lb\g^{mnp} r)(d\g_{mnp} d +24 N_{mn}\Pi_p)}\over{192(\lb\l)^2}}
+ {{(r\g_{mnp} r)(\lb\g^m d)N^{np}}\over{16(\lb\l)^3}} -
{{(r\g_{mnp} r)(\lb\g^{pqr} r) N^{mn} N_{qr}}\over{128(\lb\l)^4}}.$$

Finally, the similarity transformation 
${\cal O} \to e^{-U} {\cal O} e^{U} $ where
\eqn\simfour{U = \int dz c (B - v_m (\l\g^m)_\a \p ({{\G_n (\g^n\lb)^\a}\over{2(\l\lb)}})  - \bh\p c)}
transforms \brstg\ into
\eqn\brsth{ e^{-U} e^{-R'''} e^{-R''} e^{-R'} Q_{B-RNS-GSS} e^{R'} e^{R''} e^{R'''} e ^U =}
$$
 \int dz (\l^\a d_\a +\wtb^{\a} r_\a + \gh b+ u_m \Gb^m) - \etah_0.$$
 The usual quartet argument implies that the cohomology of \brsth\ is independent
 of 
\eqn\comind{(u_m, v^m;  \G_m, \Gb^m;  \gh, \bh; b,c ;  \lb_\a, \wtb^{\a};  r_\a, s^\a),}
 so one recovers the original pure spinor BRST operator
 $Q_{pure} =\int dz  \l^\a d_\a$ in the small Hilbert space defined with $\etah_0$.
 Note that this last similarity transformation with $U$ of \simfour\ shifts the Virasoro $b$ ghost to
 \eqn\shiftb{ e^{-U} b e^U = b + B - v_m (\l\g^m)_\a \p ({{\G_n (\g^n\lb)^\a}\over{2(\l\lb)}})  - \bh\p c - \p (\bh c).}
 
 So after twisting the spin-half $\psi^m$ variables into spin-zero and spin-one variables and performing a similarity transformation, the B-RNS-GSS BRST operator has been mapped into the pure spinor BRST operator.  Finally, after gauging away the dependence on the non-minimal variables in \comind, the unintegrated and integrated vertex operators of \app\ and \intu\ are mapped into the usual pure spinor massless vertex operators 
\eqn\usualv{V = \l^\a A_\a(x,\t), \quad U = 
 \int dz [ \p\t^\a A_\a(x,\t) + \Pi^m A_m (x,\t) + d_\a W^\a (x,\t) +\half N^{mn} F_{mn}(x,\t)].}

\vskip 10pt

{\bf Acknowledgements:}
I would like to thank
CNPq grant 311434/2020-7
and FAPESP grants 2016/01343-7, 2019/24277-8 and 2019/21281-4 for partial financial support.

\footatend\vfill\supereject\immediate\closeout\rfile\writestoppt
\baselineskip=14pt\centerline{{\bf References}}\bigskip{\frenchspacing%
\parindent=20pt\escapechar=` \input refs.tmp\vfill\eject}\nonfrenchspacing
\end